\documentclass[english,aps,prl,twocolumn,showpacs]{revtex4-1}
\usepackage{lmodern}

\usepackage[T1]{fontenc}
\usepackage[latin1]{inputenc}
\usepackage{graphicx}
\usepackage{amssymb}

\makeatletter

\usepackage{lmodern}
\usepackage{subfig}

\makeatletter

\usepackage{lmodern}

\makeatletter

\usepackage{lmodern}

\makeatletter

\usepackage{lmodern}

\makeatletter

\makeatletter

\usepackage{epsfig}

\usepackage{dcolumn}

\usepackage{bm}

\usepackage{bbm}

\usepackage{wasysym}

\newlength{\textwidthm}

\makeatother

\makeatother

\makeatother

\makeatother

\makeatother

\usepackage{babel}
\makeatother
\begin{document}

\title{Symmetry classification of energy bands in graphene}

\author{E. Kogan}
\email{Eugene.Kogan@biu.ac.il}
\affiliation{Department of Physics, Bar-Ilan University, Ramat-Gan 52900,
Israel}
\author{V. U. Nazarov}
\email{nazarov@gate.sinica.edu.tw}
\affiliation{Research Center for Applied Sciences, Academia Sinica, Taipei 11529, Taiwan}
\date{\today}

\begin{abstract}
We present the results of the first principle calculations
 of the energy bands in graphene and their symmetry classification. The valence bands and four lowest conduction bands are classified according to their
symmetry  at the  points $\Gamma$ and $K$. Merging of the bands is
interpreted  in the framework of the group theory.
\end{abstract}


\maketitle

\section{Introduction}

Since   graphene   was first isolated experimentally \cite{novoselov},  it is in the focus of attention of both   theorists and experimentalists.
Obviously, our understanding of graphene starts from  the knowledge of its energy bands.
It is  commonknowledge that the highest valence ($\pi$) band and the lowest lying conduction ($\pi^*$) band merge at the Fermi level at point $K$ (the corner of the Brillouin zone).  The  dispersion law of these two bands (obtained in the tight--binding approximation) is also well known since the seminal paper by Wallace \cite{wallace,castro}.
The other bands in graphene were also studied previously, although in a much less detailed way than the above mentioned two.  In particular, there were performed first principle calculations of the band structure of graphene
\cite{latil,wehling,trevisanutto,silkin,suzuki}.

However, a crucial element in our understanding of the band structure of graphene  is still missing, that is the symmetry analysis of the bands.
Such analysis (for any  periodic crystal  Hamiltonian) is most easily performed with the help of elementary group theory  \cite{kittel,harrison}. The symmetry of crystal is characterized by a point group $R$.  Any operation of this group
(save the unit transformation)
takes a general  wavevector ${\bf k}$ into a distinct one. However, for some special choices of ${\bf k}$
some of the operations of the crystal symmetry group will take ${\bf k}$ into itself rather  than into a distinct wavevector.
These particular operations are called the group of ${\bf k}$;  it is a
subgroup of the full symmetry group of the crystal.
Lines in the Brillouin zone for which the group of the wavevector contains elements other than the unit element are called symmetry lines.
At special points in the Brillouin zone the group of the wavevector may be larger than that on symmetry lines which thread it;   these are called symmetry points.
We may use a state (states) corresponding to such a special wavevector to generate a representation for the group of ${\bf k}$ \cite{kittel,harrison}.
(For an arbitrary wavevector, of course, the group of the wavevector is simply $E$,  and
the only irreducible representation which may be realized by the states corresponding to such a wavevector is the unit representation.)

 We may  classify  states at a wavevector corresponding to a symmetry point according to the irreducible representation
of the group of the wavevector at that point. As the wavevector then moves away from the point, the group of the wavevector becomes smaller
and some of the degeneracies will be split.We may determine the irreducible representations
into which the original representation will split. Conditions relating the irreducible representations of adjoining points and lines
are called compatibility relations and were discussed for the first time in Ref. \cite{bouckaert}. The early classical papers in the field are reprinted in Ref. \cite{knox}.
The energy band calculations are most commonly carried out along symmetry lines in the Brillouin zone.

\section{Symmetry classification of the energy bands}

We have calculated the band-structure of graphene along the
$\Gamma$--$K$--$M$--$\Gamma$ line with the code Elk \cite{Elk} which
implements the
full potential linearized plane wave method (FP-LAPW) \cite{Singh-94}
and the local-density approximation (LDA)
exchange-correlation (xc) potential.
The k-vector grid was unshifted with 64 $\times$ 64 $\times$ 1 points with
all the symmetries applied. The separation $d$ between the periodically
repeated in the perpendicular direction graphene
layers was chosen at 200 bohr. With this, the elimination of the inter-layer
interaction has been ensured.

\begin{figure}[h]
\begin{center}
\centering
\includegraphics[width=0.45\textwidth]{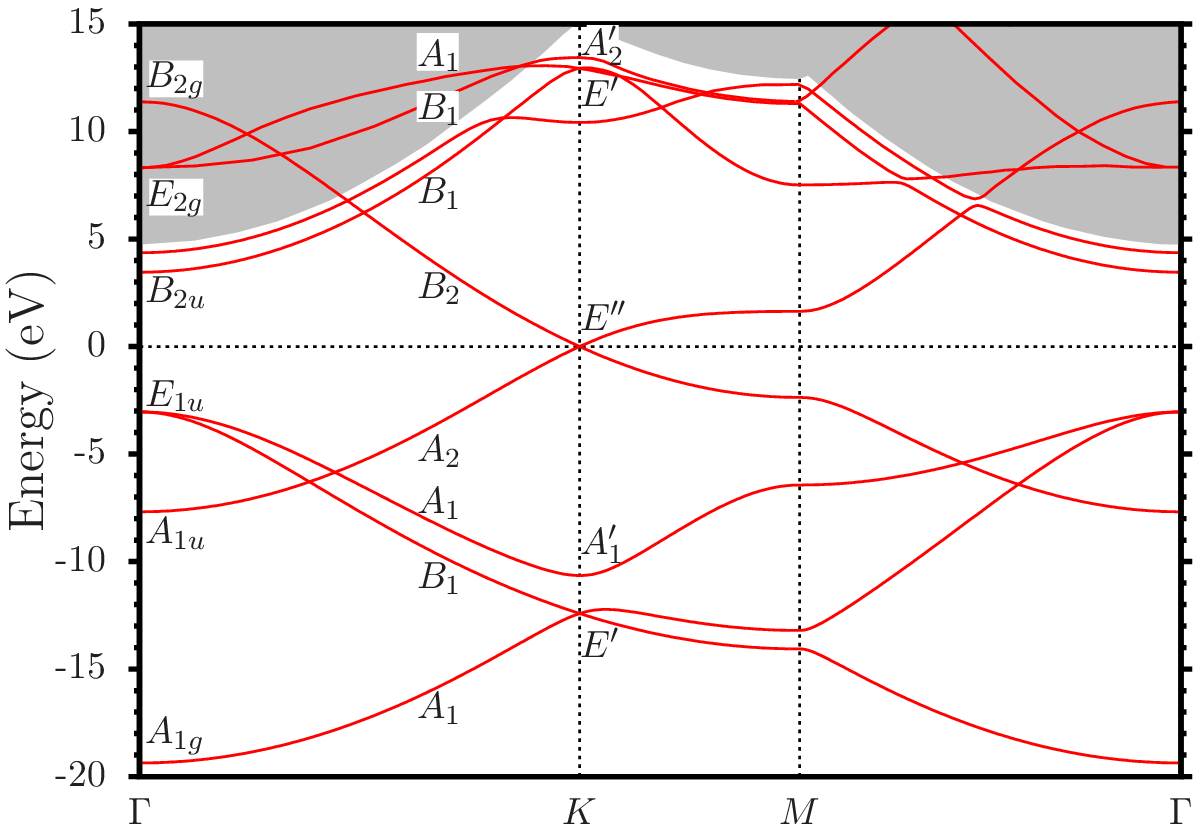}
\caption{\label{fig:bands}
(Color online) Graphene band-structure evaluated with use of the FP-LAPW
method. The red-marked lines
are well-converged single graphene layer bands, while gray background
corresponds to continuous spectrum.
}
\end{center}
\end{figure}

The results of the calculation are presented in Fig. \ref{fig:bands}. The curves
marked  have been found well converged with respect to the
further increase of $d$ and they represent the intrinsic nine lowest-lying
valence and conduction bands of graphene.
 While the overall agreement
with previous pseudopotential-based
band-structure calculations
\cite{latil,trevisanutto,wehling,Silkin-09} is found, one
important point warrants a special noting: In
Ref.~\onlinecite{Silkin-09}, the two low-lying (at $\Gamma$ point) bands
were attributed
to image-potential states. Since, in contrast to
Ref.~\onlinecite{Silkin-09}, we do not impose an additional image potential,
and since LDA for xc potential does not account for the latter
\cite{Lang-70},
we conclude that the two bands in question are ordinary Kohn-Sham LDA
energy bands rather than the image-state bands.

The gray background in Fig. \ref{fig:bands} represents the continuous spectrum
which is extrapolated in the $d\rightarrow\infty$ limit
from the quasi-continuous one obtained from the calculations.
The edge of the continuum is at approximately 4.6 eV at the $\Gamma$
point which gives the vacuum level in this calculation.
The bands
entering continuum (red lines inside the gray background)
turn into resonances, i.e., the corresponding wave-functions leak out
from the graphene plane into vacuum. However, this leakage is apparently
weak
as indicated by the stability of the corresponding bands with respect to
the $d$ increase. The studies of the width of the resonances, including those corresponding to the higher bands, will be submitted for publication separately.

The symmetry points in graphene are point $\Gamma$, the center of the Brillouin zone,  points $K$, which are corners of the Brillouin zone, and  points $M$, which are the centers of the edges of the Brillouin zone. The symmetry lines are $\Gamma$-$K$, $\Gamma$-$M$ and $K$-$M$ lines.
The small group of vector ${\bf k}$ at   point $\Gamma$ is $D_{6h}$.
The small group of vector ${\bf k}$ at   point $K$ is $D_{3h}$.
The small group of vector ${\bf k}$ at   lines $\Gamma$-$K$  is $C_{2v}$.

The representation of the group  $D_{6h}$ we can obtain  noticing that
\begin{eqnarray}
D_{6h}=D_6\times C_i.
\end{eqnarray}
Thus  representation, say, $A_1$ of group $D_6$ begets two representations of the group $D_{6h}$: $A_{1g}$ and $A_{1u}$, where the letter
$g$ or $u$ means that the representation is  even or odd with respect to inversion respectively.
The $z$-axis is always chosen perpendicular to the graphene plane.
When considering the $\Gamma$-$K$ line,  the $x$ axis is chosen in the direction of the line.
\begin{table}
\begin{tabular}{|l|l|l|cccccc|}
\hline
$D_6$ & &  & $E$ & $C_2$ & $2C_3$ & $2C_6$ & $3U_2$ & $3U_2'$  \\
 & $C_{6v}$ &  & $E$ & $C_2$ & $2C_3$ & $2C_6$ & $3\sigma_v$ & $3\sigma_v'$ \\
 & & $D_{3h}$  & $E$ & $\sigma_h$ & $2C_3$ & $2S_3$ & $3U_2$ & $3\sigma_v$ \\
\hline
$A_1$ & $A_{1}$ & $A_1'$ & 1 & 1 & 1 & 1 & 1 & 1 \\
$A_{2};z$ & $A_{2}$ & $A_2'$ & 1 & 1 & 1 & 1 & -1 & -1 \\
$B_1$ & $B_{2}$ & $A_1''$ & 1 & -1 & 1 & -1 & 1 & -1 \\
$B_2$ & $B_{1}$ & $A_2''$ & 1 & -1 & 1 & -1 & -1 & 1 \\
$E_2$ & $E_{2}$ &$E'$ & 2 & 2 & -1 & -1 & 0 & 0 \\
$E_{1}$  & $E_{1}$  & $E''$ &2 & -2  & -1 & 1 & 0 & 0 \\
\hline
\end{tabular}
\caption{Character tables }
\label{table:d66}
\end{table}

\begin{table}
\begin{tabular}{|c|cccc|}
\hline
$C_{2v}$  & $E$ & $C_2$ & $\sigma_v^{xy}$ & $\sigma_v^{xz}$ \\
\hline
$A_1$    & 1 & 1 & 1 & 1  \\
$A_2$ &  1 & 1 & -1 & -1  \\
$B_1$ &  1 & -1 & 1 & -1  \\
$B_2$ &  1 & -1 & -1 & 1  \\
\hline
\end{tabular}
\begin{tabular}{|c|ccc|}
\hline
$C_{3v}$  & $E$ & $2C_3$ & $3\sigma_v$ \\
\hline
$A_1$    & 1 & 1 & 1   \\
$A_2$ &  1 & 1 & -1   \\
$E$ & 2 & -1 & 0   \\
\hline
\end{tabular}
\begin{tabular}{|c|ccc|}
\hline
$C_{3}$  & $E$ & $C_3$ & $C_3^2$ \\
\hline
$A$    & 1 & 1 & 1   \\
$E;x+iy$ &  1 & $\epsilon$ & $\epsilon^2$   \\
$E;x-iy$ & 1 & $\epsilon^2$ & $\epsilon$   \\
\hline
\end{tabular}
\caption{Character tables; $\epsilon=e^{2\pi i/3}$}
\label{table:d2}
\end{table}

The compatibility relations of the groups are obtained by taking into account that irreducible representations of the groups $D_{6h}$ and $D_{3h}$
become reducible when we consider them as the representations of the group $C_{2v}$. The expansion of these representations with respect to irreducible representations of the group $C_{2v}$ can be easily done using the tables of characters (Tables \ref{table:d66} and \ref{table:d2}).
Equation
\begin{eqnarray}
\label{ex}
a^{(\alpha)}=\frac{1}{g}\sum_G\chi(G)\chi^{(\alpha)}(G)
\end{eqnarray}
shows how many times a given irreducible representation is contained in a reducible one.
All the relevant compatibility  relations are  presented in Table \ref{table:c1}.
\begin{table}
\begin{tabular}{|c|c|c|c|}
	\hline
$C_{2v}$ & $D_{6h}$ & $D_{3h}$ & $D_{2h}$ \\
\hline
Rep &\multicolumn{3}{c|} {Compatible with}  \\
	\hline
$A_{1}$   & $A_{1g},B_{1u},E_{1u},E_{2g}$ & $A_1',E'$ & $A_g, B_{1g}, B_{2u},B_{3u}$ \\
$A_{2}$  & $A_{1u},B_{1g},E_{1g},E_{2u}$ & $A_1'',E''$ & $A_u,B_{1u},B_{2g}, B_{3g}$  \\
$B_{1}$  & $A_{2g},B_{2u},E_{1u},E_{2g}$ & $A_2',E'$ &  $A_g,B_{1g},B_{2u},B_{3u}$ \\
$B_{2}$  & $A_{2u},B_{2g},E_{1g},E_{2u}$ & $A_2'',E''$ & $A_u,B_{1u},B_{2g},B_{3g}$ \\
\hline
\end{tabular}
\caption{Compatibility relations for the honeycomb lattice ($\Gamma-K$ line)}
\label{table:c1}
\end{table}

Now let us come directly to the classification of the energy bands.
We assume that the representation realized by the lowest valence band at  point $\Gamma$ is the maximum symmetry representation $A_{1g}$. From  Table \ref{table:c1}
we immediately come to the conclusion that the band at the line $\Gamma-K$ is characterized by the representation $A_1$.
 The representation $A_1$ is compatible with
one-dimensional representation $A_1'$ and two-dimensional representation $E'$ at the point $K$. But because of degeneracy in this point we come to the conclusion that the lowest valence band realizes
the representation $E'$  at  point $K$, and the other band, merging with it at  point $K$, at  line $\Gamma$-$K$ realizes the representation $B_1$.
Hence the third band, merging with the second one at the point $K$, at  line $\Gamma$-$K$ realizes the representation $A_1$.
The second and the third band realize at  point $\Gamma$ either the representation   $E_{1u}$ or $E_{2g}$.
Assuming that these two bands  produce a bonding (that is symmetric) orbital, we come to the conclusion that they realize at the point $\Gamma$ representation $E_{1u}$.
Taking into account that $\sigma_h=IC_2$, we deduce that these valence bands are even with respect to the $\sigma_h$ operation.

Let us come to two bands which merge at the Fermi level at the point $K$. It is  common knowledge that they are antisymmetric  with respect to the $\sigma_h$ operation (see  Sec. \ref{tb}). Among the two bands in question the valence band ($\pi$-band) is characterized at  point $\Gamma$ by the representation $A_{1u}$ (bonding and, hence, symmetric orbital), and  the conduction band ($\pi^*$-band) is characterized at  point $\Gamma$ by the representation $B_{2g}$ (antibonding and, hence, antisymmetric orbital).
At  line $\Gamma$-$K$ they are characterized by the representations $A_2$ and $B_2$ respectively, and at  point $K$ they are characterized by the representation $E''$ \cite{slon,thomsen}.
It was mentioned already in the classical paper \cite{slon}, that a two dimensional representation realized by the wave functions at  point $K$ means that the dispersion law in the vicinity of this point can be presented by a circular cone.
Similar reasoning can be applied to  the lowest lying conduction bands.
All the  results of symmetry classification are presented on Fig. \ref{fig:bands}.

The most prominent features in the band structure presented in Fig. 1 is the merging of the bands. In  the next Section we will interpret merging of the $\pi$ and $\pi^*$ bands at  point $K$ and merging of the $\sigma$ bands at  point $\Gamma$ using the tight binding model. In  the Section after the next one we will
interpret  merging of the  bands at  point $K$ using the quasi--free electrons model.

\section{Tight--binding Hamiltonian + group theory}
\label{tb}

It is interesting to look at the proposed symmetry classification of the bands in general and at their merging in particular from the point of view of the tight--binding model.
In isolated form, carbon has six electrons in the orbital configuration
$1s^{2}2s^{2}2p^{2}$. When arranged in the honeycomb crystal, two electrons remain in the core
$1s$ orbital, and are traditionally ignored  in band calculations. The remaining four electrons occupy four valence bands; there are also conduction bands, of which a few lowest ones are of  interest.
The tight--binding Hamiltonian for $sp$-bonded systems includes  four orbitals per atom: $s,p_x,p_y,p_z$. The Hamiltonian being symmetric with respect to reflection in the graphene plane,
the bands built from the $p_z$ orbitals decouple from those built from $s,p_x,p_y$ orbitals. The former are odd with respect to reflection in the graphene plane, the latter are even.
The classification with respect to reflection in the graphene plane already done, we can talk about $C_{6v}$ and $C_{3v}$ groups instead of $D_{6h}$ and $D_{3h}$.

\subsection{The $\pi$ bands}

The structure of graphene
can be seen as a triangular lattice with a basis of two
atoms per unit cell, displaced from each other by any one (fixed) vector connecting two sites of different sub-lattices, say
${\bf \delta}=-a\left(1,0\right)$.
The general  Hamiltonian for the $\pi$ bands is
\begin{eqnarray}
\label{ham}
H =-
\left(\begin{array}{cc}
\sum_{\bf a} t'({\bf a})e^{i{\bf k\cdot a}} & \sum_{\bf a}t({\bf a}+{\bf \delta})e^{i{\bf k\cdot}({\bf a}+{\bf \delta})}\\
\sum_{\bf a}t^*({\bf a}+{\bf \delta})e^{-i{\bf k\cdot}({\bf a}+{\bf \delta})} &  \sum_{\bf a}t'({\bf a})e^{i{\bf k\cdot a}} \end{array}\right),\nonumber\\
\end{eqnarray}
where ${\bf a}$ is an arbitrary  lattice vector, that is a linear combination of
${\bf a}_1=\frac{a}{2}\left(3,\sqrt{3}\right)$, ${\bf a}_2=\frac{a}{2}\left(3,-\sqrt{3}\right)$.

The selection rule for matrix elements
gives
\begin{eqnarray}
\label{zero}
\sum_{\bf a}t({\bf a}+{\bf \delta})e^{i{\bf K\cdot}({\bf a}+{\bf \delta})}=0,
\end{eqnarray}
where {\bf K} is a  corner of the Brillouin zone.
In fact, we are dealing with the product of two functions: $t({\bf a}+{\bf \delta})$ realizes the unit representation
of the point symmetry group $C_{3}$ (the full symmetry group of the inter--sublattice hopping is $C_{3v}$, but the restricted symmetry $C_3$  will be enough to prove the cancellation).
As far as the function $e^{i{\bf K\cdot}({\bf a}+{\bf \delta})}$ is concerned, rotation of the lattice by the angle $2\pi/3$,
say anticlockwise,
 is equivalent to rotation of the vector ${\bf K}$ in the opposite direction,
 that is to substitution of the three equivalent corners of the Brillouin zone: ${\bf K}_1\to {\bf K}_2\to{\bf K}_3\to {\bf K}_1$, where
${\bf K}_1=\left(\frac{2\pi}{3a},\frac{2\pi}{3\sqrt{3}a}\right)$, ${\bf K}_2=\left(0,-\frac{4\pi}{3\sqrt{3}a}\right)$
and ${\bf K}_3=\left(-\frac{2\pi}{3a},\frac{2\pi}{3\sqrt{3}a}\right)$. Thus due to the rotation $e^{i{\bf K\cdot}({\bf a}+{\bf \delta})}$
is multiplied by the factor $\epsilon^2$ and
realizes  $x-iy$ representation of the group $C_{3}$. Because each of the multipliers in Eq. (\ref{zero}) realizes different irreducible representation of the symmetry group, the matrix element is equal to zero.
Simply speaking, at a point {\bf K} the sublattices become decoupled, and this explains the degeneracy of the electron states in this point (these points) or, in other words, merging of the two branches of the single Brillouin zone.

On the other hand, generally
\begin{eqnarray}
\sum_{\bf a} t'({\bf a})e^{i{\bf K\cdot a}}\neq 0.
\end{eqnarray}
To understand this statement  consider the maximum symmetry group of the intra--lattice hopping: $C_{6v}$.
The function $t'({\bf a})$ realizes the $A_1$ representation
of the  group.
Applying  Eq. (\ref{ex}) we see that reducible representation   of the group $C_{6v}$,  realized by the two functions $e^{i{\bf K\cdot a}}$ and $e^{i{\bf K'\cdot a}}$ can be decomposed as $A_1+B_2$.

In addition, the tight--binding model provides us with a simple explanation of why the dispersion law in the vicinity of the merging points is linear, that is why these points are Dirac points.
The dispersion law for the Hamiltonian (\ref{ham}) is given by equation
\begin{eqnarray}
\label{dispersion}
F(E,{\bf k})=0,
\end{eqnarray}
where
\begin{eqnarray}
\label{dispersion2}
F(E,{\bf k})=\left|\begin{array}{cc}
E+\sum_{\bf a} t'({\bf a})e^{i{\bf k\cdot a}} & \sum_{\bf a}t({\bf a}+{\bf \delta})e^{i{\bf k\cdot}({\bf a}+{\bf \delta})}\\
\sum_{\bf a}t^*({\bf a}+{\bf \delta})e^{-i{\bf k\cdot}({\bf a}+{\bf \delta})} & E+ \sum_{\bf a}t'({\bf a})e^{i{\bf k\cdot a}} \end{array}\right|.\nonumber\\
\end{eqnarray}

In mathematics
the Dirac points, we are dealing with, are called conical points of the surface; if the surface is given by Eq. (\ref{dispersion}),
the conditions for the conical points are \cite{goursat}
\begin{eqnarray}
\label{dirac}
\frac{\partial F}{\partial E}=0,\qquad
 \frac{\partial F}{\partial {\bf k}}=0.
\end{eqnarray}
Recalling the rule for  differentiating of a determinant, we realize that Eq. (\ref{zero}) guaranties that the conditions (\ref{dirac}) for ${\bf k}={\bf K}$.
This explains linearity of the spectrum in the vicinity of the points {\bf K}({\bf K}').

\subsection{The $\sigma$ bands}

We will concentrate upon the dispersion law of the $\sigma$ bands along the $\Gamma$-$K$ line. Mirror reflection in this line
exchanges the graphene sublattices, changes $y$ to $-y$, and does not change the wave vector. It means that the valence bands
labeled by $A_1$ are constructed from sub--lattice symmetric combinations of $s$ and $p_x$ orbitals and from
sub--lattice antisymmetric combinations of   $p_y$ orbitals. The valence band
labeled by $B_1$ is constructed from sub--lattice antisymmetric combinations of $s$ and $p_x$ orbitals and from
sub--lattice symmetric combinations of   $p_y$ orbitals.

We can be more specific speaking about  the valence states at  points $\Gamma$ and $K$. The states with the symmetry $A_{1g}$ and $A_1'$ are the sub--lattice symmetric combinations of $s$  orbitals.
The states with the symmetry $E_{1u}$ and $E'$ are the (degenerate) sub--lattice symmetric combinations of $p_x$ and $p_y$ orbitals.

Looking at the conduction bands we realize that the two bands which merge at the point $\Gamma$ with the symmetry $E_{2g}$
are probably of the same nature as the two valence bands which merge at the point $\Gamma$ with the symmetry $E_{1u}$,
and the states with the symmetry $E_{2g}$ are the (degenerate) sub--lattice antisymmetric combinations of $p_x$ and $p_y$ orbitals.
(In the language of quantum chemistry sub--lattice symmetric and antisymmetric combinations are bonding and antibonding orbitals respectively.)

\section{Quasi--free electrons + group theory}
\label{qf}

If we consider separately $\sigma$ and $\pi$ states we can
ignore the third dimension and treat  graphene as purely two dimensional.
The Hamiltonian of the electrons is
\begin{eqnarray}
H=H_0+V,
\end{eqnarray}
where $H_0$ is the Hamiltonian of the free electrons and $V$ is the crystalline potential.
The following reasoning  uses the extended Brillouin zone scheme \cite{madelung} and is
 the application of group theory to quantum mechanics \cite{heine}.

All the  eigenstates of the Hamiltonian $H_0$ with the same $|{\bf k}|$ are degenerate. The potential $V$ having lower symmetry
partially (nearly completely)  removes the degeneracy. To find the maximum splitting we  should find
among the continuum of the vectors with the given value of $|{\bf k}|$ those which realize an irreducible representation of this or that
sub--group of the space group of the crystal. For arbitrary value of $|{\bf k}|$ each state
is symmetric only with respect to the sub--group of translations and
realizes a one--dimensional representation of this sub--group.

However,  the states with the wavevector ${\bf K}+{\bf b}$, where ${\bf b}$ is an arbitrary vector of the inverse lattice (corners of the Brillouin zones)
have higher symmetry. They
realize a representation (infinite dimensional) of the group which includes in addition to translations the point group $C_{3v}$.
We will expand this representation into the irreducible ones in two stages.

First we will divide all the vectors ${\bf K}+{\bf b}$
into the sets such, that elements of a given set are connected with each other by the operations of the group $C_{3v}$.
Such sets turn our to be triplets.
For example, the states with the wavevectors
${\bf K}_1,{\bf K}_1-{\bf b}_1,{\bf K}_1-{\bf b}_1-{\bf b}_2$, where ${\bf b}_1=\frac{2\pi}{3a}\left(1,\sqrt{3}\right)$ and
${\bf b}_2=\frac{2\pi}{3a}\left(1,-\sqrt{3}\right)$ are the elementary reciprocal-lattice vectors.
Another example is the states having wavevectors
${\bf K}_1-{\bf b}_2,{\bf K}_1+{\bf b}_2,{\bf K}_1-2{\bf b}_1-{\bf b}_2$. Still another example is two triplets having the same energy:
${\bf K}_1-2{\bf b}_1-2{\bf b},{\bf K}_1+{\bf b}_1,{\bf K}_1-{\bf b}_1+{\bf b}_2$ and ${\bf K}_1-{\bf b}_1-2{\bf b}_2,{\bf K}_1-2{\bf b}_1,
{\bf K}_1+{\bf b}_1+{\bf b}_2$.

At the second stage, applying Eq. (\ref{ex}) we  decompose  the  reducible representation $R$ of the symmetry group $C_{3v}$, generated by a triplet into  the irreducible representations. The triplet generates natural representation of the group $C_{3v}$ \cite{riley}.
\begin{table}
\begin{tabular}{|c|ccc|}
\hline
$C_{3v}$  & $E$ & $2C_3$ & $3\sigma_v$ \\
\hline
$R$    & 3 & 0 & 1  \\
\hline
\end{tabular}
\caption{Characters of natural representation}
\end{table}
Applying Eq. (\ref{ex}) we obtain
\begin{eqnarray}
\label{simple}
R=A_1+E.
\end{eqnarray}
It should be noted that the result does not depend on $V$ being in any sense small \cite{heine}.

Recalling the symmetry of the states relative to reflection in the plane of graphene, Eq. (\ref{simple}) can be rewritten as
\begin{eqnarray}
R=A_1'+E'
\end{eqnarray}
for $\sigma$ states, and as
\begin{eqnarray}
R=A_2''+E''
\end{eqnarray}
for $\pi$ states.

A  picture that emerges becomes clear in  the extended Brillouin zone scheme, where the Brillouin zones
fill the whole plane. Each Brillouin zone being a hexagon,  three zones meet at their corners;
two of them  merge.

To find the dispersion law in the vicinity of the merging point  we may consider a reduced Hamiltonian,
\begin{eqnarray}
\label{ham5}
H =
\left(\begin{array}{cc}
H_{11} & H_{12}\\
H_{21} &  H_{22} \end{array}\right),
\end{eqnarray}
where each element is a linear function of $k_x,k_y$ (${\bf k}$ is the deviation of the wavevector from the merging point), and we have shifted the energy
axis, such that the energy in the merging point is now zero.
The dispersion law for the Hamiltonian (\ref{ham5}) is given by equation
\begin{eqnarray}
E=c_ik_i\pm \sqrt{d_{ij}k_ik_j},
\end{eqnarray}
describing two cones, which is a general situation for levels crossing \cite{landau}.
However, in the case considered, we can be more specific about these cones.
Because  any vector in the $k_x,k_y$ plane compatible with the symmetry $C_{3v}$ is identically equal to zero, and any  tensor of rank two
 compatible with the symmetry  is proportional to the unity tensor, the dispersion law is just
\begin{eqnarray}
E\sim\pm |{\bf k}|,
\end{eqnarray}
and the cones are circular, with the  axis  perpendicular to the $k_x,k_y$ plane.

The important role played by discrete symmetries in protecting
a $k$--linear dispersion in graphene was pointed out  in
Ref. \cite{manes}.
Ref. \cite{malard} was an important precedent in applying group theory methods to graphene.
The appearance of massless Dirac fermions  under conditions of hexagonal symmetry was considered in Ref. \cite{park}.
Group theory was
used to derive an invariant expansion of the Hamiltonian for electron states near the K points of the graphene
Brillouin zone  in Ref. \cite{winkler}.
The influence of stress on the bands merging was analyzed in Refs. \cite{pereira,ribeiro}.
The influence of spin-orbit interactions on the band structure of graphene was studied in Ref.
\cite{min}.

\section{Acknowledgements}

The work was finalized during one of the authors (E.K.) visit to DIPC, San Sebastian.
The authors are grateful to V. Silkin for very useful discussion. The other author
(V.U.N.) acknowledges partial support from National Science Council, Taiwan, Grant No. 100-2112-M-001-025-MY3.


\begin{thebibliography}{99}

\bibitem{novoselov} K. S. Novoselov, A. K. Geim, S. V. Morozov, D. Jiang,
Y. Zhang, S. V. Dubonos, I. V. Grigorieva, A. A. Firsov, Science {\bf 306}, 5696 (2004).

\bibitem{wallace} P.R. Wallace, Phys. Rev. {\bf 71}, 622 (1947).

\bibitem{castro} A. H. Castro Neto, F. Guinea, N. M. R. Peres, K. S. Novoselov and A. K. Geim,
Rev. Mod. Phys. {\bf 81}, 109 (2009).

\bibitem{latil} S. Latil and L. Henrard, Phys. Rev. Lett. {\bf 97}, 036803 (2006).

\bibitem{wehling} T. O. Wehling, I. Grigorenko, A. I. Lichtenstein, and A. V. Balatsky, \prl {\bf 101}, 216803 (2008).

\bibitem{trevisanutto} P. E. Trevisanutto, C. Giorgetti, L. Reining, M. Ladisa, V. Olevano, Phys.
Rev. Lett. {\bf 101}, 226405 (2008).

\bibitem{silkin} V. M. Silkin, J. Zhao, F. Guinea, E. V. Chulkov, P. M. Echenique, and H. Petek, \prb {\bf 80}, 121408 (2009).

\bibitem{suzuki} T. Suzuki, Y. Yokomizo, Physica E {\bf 42}, 2820 (2010).


\bibitem{kittel} C. Kittel, {\it Quantum Theory of Solids}, (John Wiley \& and Sons. Inc. New--York, London 1963).

\bibitem{harrison} W. A. Harrison, {\it Solid State Theory}, (McGraw Hill Book Company, New York--London--Toronto, 1970).

\bibitem{bouckaert} L. P. Bouckaert, R. Smoluchowski and E. Wigner, Phys. Rev. {\bf 50}, 58 (1936).

\bibitem{knox} R. S. Knox and S. Gold, {\it Symmmetry in the Solid State}, (W. A. Benjamin, Inc., 1964, New York, Amsterdam).

\bibitem{Elk} http://elk.sourceforge.net, 2011.

\bibitem{Singh-94} D. H. Singh,
{\it Planewaves, Pseudopotentials, and the LAPW method},
(Kluwer, Boston, 1994).


\bibitem{Silkin-09} V. M. Silkin, J. Zhao, F. Guinea, E. V. Chulkov,
P. M. Echenique,  and H. Petek,
Phys. Rev. B {\bf 80}, 121408 (2009).

\bibitem{Lang-70} N. D. Lang  and W. Kohn, Phys. Rev. B, {\bf 1}, 4555 (1970).


\bibitem{slon}J. C. Slonczewski and P. R. Weiss,  Phys. Rev. {\bf 109}, 272 (1958).

\bibitem{thomsen}  C. Thomsen,  S. Reich,  J. Maultzsch, {\it Carbon Nanotubes: Basic Concepts and Physical Properties}, (Wiley Online Library,
2004 WILEY-VCH Verlag GmbH).


\bibitem{goursat} E. Goursat, {\it Course D'Analyse Mathematique} [cited by Russian translation: {\it Kurs Matematicheskogo Analisa}, (Moskow 1933), Vol. I, Part 1, p 100.]


\bibitem{heine} V. Heine, {\it Group Theory in Quantum Mechanics} (Pergamon Press, 1970).

\bibitem{riley} K. F. Riley, M. P. Hobson, and S. J. Bence, {\it Mathematical Methods for Physics and Engeneering, 3ed.} (Cambridge University Press, 2006).

\bibitem{madelung} O. Madelung, {\it Introduction to Solid--State Theory} (Springer--Verlag Berlin Heidelberg New York, 1996).

\bibitem{landau} L. D. Landau and E. M. Lifshitz, {\it Quantum Mechanics}, (Pergamon Press, 1991).

\bibitem{manes} J. L. Manes, F. Guinea, and M. A. H. Vozmediano, Phys. Rev. B
{\bf 75}, 155424 (2007);  J. L. Manes, arXiv:1109.2581.

\bibitem{malard} L. M. Malard, M. H. D. Guimaraes, D. L. Mafra, M. S. C. Mazzoni, and A. Jorio, \prb {\bf 79}, 125426 (2009).

\bibitem{park} C.-H. Park and S. G. Louie, Nano Lett. {\bf 9}, 1793 (2009).

\bibitem{winkler} R. Winkler, U. Zulicke, Phys. Rev. B {\bf 82}, 245313 (2010).

\bibitem{pereira} V. M. Pereira,  A. H. Castro Neto and N. M. R. Peres, Phys. Rev. B 80, 045401 (2009).

\bibitem{ribeiro} R. M. Ribeiro, V. M Pereira, N. M. R. Peres, P. R. Briddon, and A. H. Castro Neto,  New J. Phys. {\bf 11}, 115002 (2009).

\bibitem{min} H. Min, J. E. Hill, N. A. Sinitsyn, B. R. Sahu, Leonard Kleinman, and A. H. MacDonald,  Phys. Rev. B {\bf 74}, 165310 (2006).


\end{thebibliography}
\end{document}